\documentclass[runningheads]{llncs}
\usepackage{graphicx}
\usepackage{lipsum}

\newcommand\blfootnote[1]{%
  \begingroup
  \renewcommand\thefootnote{}\footnote{#1}%
  \addtocounter{footnote}{-1}%
  \endgroup
}

\graphicspath{ {media/} }
\begin{document}
\title{Knowledge Distillation for Brain Tumor Segmentation}
%
%
\author{Dmitrii Lachinov\inst{1}\orcidID{0000-0002-2880-2887} \and
Elena Shipunova\inst{2} \and
Vadim Turlapov\inst{3}\orcidID{0000-0001-8484-0565}}
\authorrunning{D. Lachinov et al.}
%
\institute{Dept. of Ophthalmology and Optometry, Medical University of Vienna\\
\email{dmitrii.lachinov@meduniwien.ac.at}\\
\and
Intel, Russian Federation\\
\email{elena.shipunova@intel.com}\\
\and
Lobachevsky State University\\
\email{vadim.turlapov@itmm.unn.ru}}
\maketitle              
\begin{abstract}
The segmentation of brain tumors in multimodal MRIs is one of the most challenging tasks in medical image analysis. The recent state of the art algorithms solving this task are based on machine learning approaches and deep learning in particular. The amount of data used for training such models and its variability is a keystone for building an algorithm with high representation power.\newline
In this paper, we study the relationship between the performance of the model and the amount of data employed during the training process. On the example of brain tumor segmentation challenge, we compare the model trained with labeled data provided by challenge organizers, and the same model trained in omni-supervised manner using additional unlabeled data annotated with the ensemble of heterogeneous models.\newline
As a result, a single model trained with additional data achieves performance close to the ensemble of multiple models and outperforms individual methods.

\keywords{BraTS  \and segmentation \and knowledge distillation \and deep learning}
\end{abstract}

\section{Introduction}
Brain tumor segmentation is a reliable instrument for disease monitoring. Moreover, it is a central and informative tool for planning further treatment and assessing the way the disease progresses. However, manual segmentation is a tedious and time-consuming procedure that requires a lot of attention from the grader.\blfootnote{\url{https://github.com/lachinov/brats2019}}\newline
To simplify clinicians workload, many automatic segmentation algorithms have been proposed recently. The majority of them utilize machine learning techniques and deep learning in particular. The downside of these approaches is the amount of data required to successfully train a deep learning model. To capture all possible variations of biological shapes, it is desired to have them present in the manually labeled dataset.\newline
The dataset provided in the scope of Brain Tumor Segmentation Challenge (BraTS 2019) \cite{dataset1,dataset2,dataset3} is the largest publicly available dataset \cite{data1,data2} with MRI scans of brain tumors. This year, it contains 259 High-Grade Gliomas (HGG) and 76 Low-Grade Gliomas (LGG) in the training set. Each MRI scan describes native (T1), post-contrast T1-weighted (T1Gd), T2-weighted (T2) and T2 Fluid Attenuated Inversion Recovery (T2-FLAIR) volumes. These images were obtained using different scanner and protocols from multiple institutions. Each scan is registered to the same anatomical template, skull stripped and resampled to the isotopic resolution. All the images were manually segmented by multiple graders who followed the same annotation protocol. The GD-enhancing tumor, the peritumoral edema, the necrotic and non-enhancing tumor core were annotated on the scans.\newline
The BraTS dataset is considered to be one of the largest publicly available medical dataset with 3D data. Despite the dataset size, it is still considered small compared to the natural images datasets, that may contain millions of samples. Datasets of such scale cover broad number of training examples, allowing algorithms to catch details that are not present in smaller size datasets. We think that the limited amount of tumor shapes and locations is the main reason why regularization techniques \cite{myronenko} work especially well in the provided scope. In this study, we are trying to solve the problem of limited dataset size from a different angle. We propose to utilize all available unlabeled data in the training process using the knowledge distillation \cite{distill1,distill2}.\newline
Originally the knowledge distillation was proposed by Hinton et. al. \cite{distill1} for transferring the knowledge of an ensemble to the single neural network. Authors introduced the new type of ensemble consisting of several big models and many specialist models which learned to distinguish fine-grained classes that the 'big' models were misclassifying. The soft labels could be utilized as a regularization during the training of the final model. The effectiveness of the proposed approach was demonstrated on MNIST dataset as well as on the speech recognition task.\newline
The variation of knowledge distillation called data distillation was introduced by Radosavovic et. al. \cite{distill2}, where authors investigated omni-supervised learning, a special case of semi-supervised learning with available labeled data as well as unlabeled internet-scale data sources. Authors proposed to annotate unlabeled data with a single model by averaging predictions produced by differently transformed input data. The automatically annotated data was then used for training of a student model on the combined dataset. The data distillation was demonstrated on the example of human keypoint detection and general object detection, where student models outperform models trained solely on labeled data.\newline
Inspired by previous papers, in this study we employ the knowledge distillation method to train student model with labeled data from BraTS 2019 challenge, automatically labeled data from BRaTS 2016 and unlabeled data from BraTS 2019 and 2018. To provide annotation for unlabeled datasets we train an ensemble of models.\newline

\section{Building an Ensemble}
The key idea of our approach is to enhance the generalization power of the student model by training it on the larger scale dataset. We achieve this by utilizing the unlabeled data available in the scope of the BraTS challenge.\newline
Starting from 2017, training dataset of the challenge includes manually annotated data, as well as data from 2012 and 2013 challenges, that was also graded manually. However, data from 2014-2016 challenges was segmented by the fusion of best-performing methods from the challenges of previous years. Later, that data was discarded. Even though the quality of annotation grew significantly, the number of available samples is still incomparable to the scale of natural images datasets. In our opinion, the quantity of training samples is as important as the quality of annotation. And at the current moment, the data we have doesn't represent all possible variations of tumor shapes and structures.\newline
During the last year's challenge, Isensee et. al. \cite{isensee2018} demonstrated that the accuracy of the model benefits from adding data from 2014-2016 challenges. We believe that this is one of the reasons why the mentioned method was ranked high during the 2018 competition. To further investigate the effectiveness of adding additional data into the training, we propose to utilize data from 2014-2016 and 2018 challenges. More specifically, we employ BraTS 2016 dataset, that is segmented automatically by the organizers. We also utilize this year's validation dataset, as well as 2018 testing dataset. Since the ground truth is unavailable, we first train an ensemble of the models. In the ensemble, we include multiple architectures \cite{myronenko,isensee2018,lachinov1} that demonstrated their performance during previous challenges. Then we automatically annotate unlabeled data with this ensemble trained solely on the BraTS 2019 training dataset; and we train a single model on the extended dataset. We describe the baseline methods that form our ensemble in the following section.\newline

\subsection{No New Net}
The first method we employ in our ensemble is UNet \cite{unet} with 3D convolutions \cite{3dunet} and minor modifications employed by Isensee et. al \cite{isensee2018} for participation in BraTS 2018 challenge, where this method was ranked second.\newline
\begin{figure}
\includegraphics[width=\textwidth]{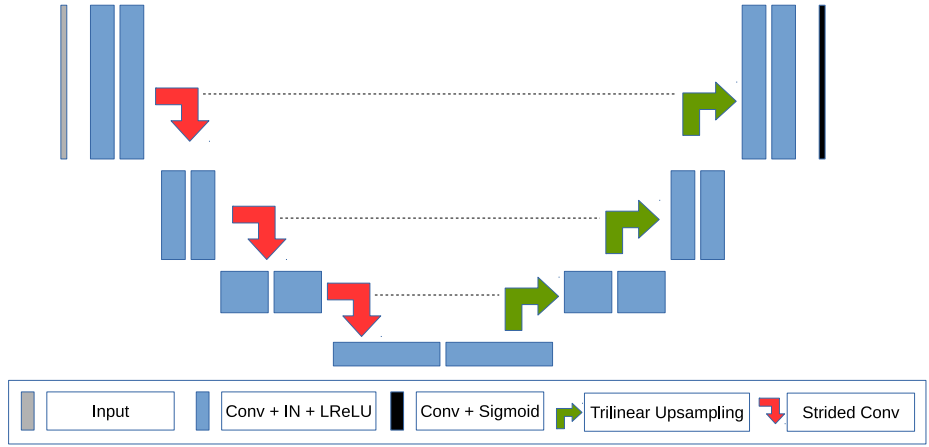}
\caption{The architecture of UNet that we use in our experiments. Dotted lines represent skip-connections. Instead of max pooling we use strided convolutions.}
\label{fig:unet}
\end{figure}
UNet is a fully convolutional decoder-encoder network with skip connections, that allows to segment fine structures well. It is especially effective in segmentation of biomedical imaging data. With slight modification its capable of showing state of the art results. For instance, Isensee et. al. replaced ReLU activations with its leaky variant with negative slope equals to $10^{-2}$. At the same time, trilinear upsampling was used in the decoder with prior filter number reduction. Instance Normalization \cite{instancenorm} was used instead of Batch Normalization \cite{batchnorm} due to inaccurate statistics estimation with small batch sizes. The network architecture is illustrated on Fig.~\ref{fig:unet}.\newline
We train this network with a patch size of 128x128x128 voxels and batch size of 2 for 160k iterations of weight updates. The initial learning rate was set to $10^{-4}$ and then dropped by a factor of 10 at 120k iterations. As in the paper \cite{isensee2018} we employed a mixture of Soft Dice Loss and Binary Cross-Entropy for training. Similarly, instead of predicting 4 target classes, we segment three overlapping regions: Whole Tumor, Tumor Core and Enhancing Tumor.
$$L^{dice}(g,p)=\frac{1}{K}\cdot\sum_{k=1}^{K}{\frac{{}2\sum{p_k\cdot g_k}}{\sum{p_k^2 + g_k^2}}}$$
$$L^{bce}(g,p) = -\frac{1}{N}\cdot\sum_{k=1}^{K}{\sum{(g_k\cdot log(p_k) + (1-g_k)\cdot log(1-p_k))}}$$\newline
where $N$ - is a number of voxels in the output, $K$ - number of classes, $g_k$ and $p_k$ is ground truth and predicted probabilities of class $k$ respectively. The resulting loss function is calculated as a sum of Soft Dice loss and BCE loss: $$L^{final}(g,p) = L^{dice}(g,p) + L^{bce}(g,p)$$

\subsection{UNet with residual connections}
\label{sec:resunet}
The second method we use in the ensemble is a UNet\cite{unet,3dunet} with residual connections \cite{resnet}. This model coupled with the autoencoder branch that imposes additional regularization on the encoder was employed by A. Myronenko \cite{myronenko}, who took the first place at previous year's challenge. \newline 
\begin{figure}
\includegraphics[width=\textwidth]{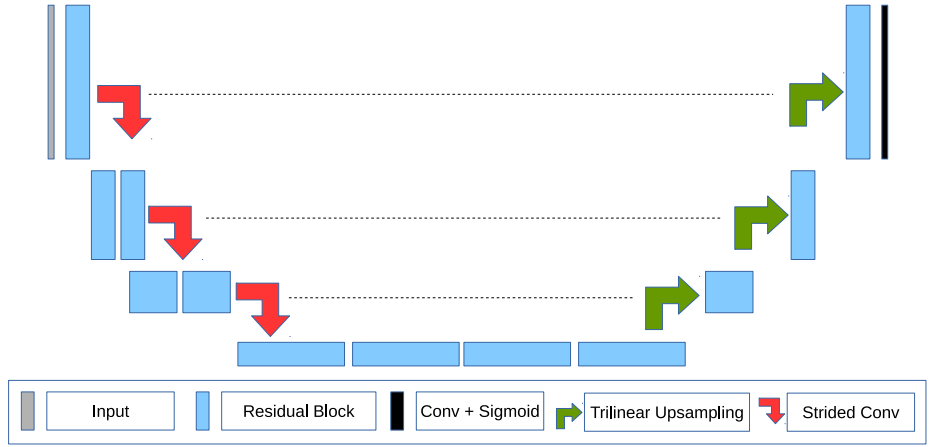}
\caption{UNet with residual blocks. Dotted arrows represent skip-connections.}
\label{fig:resunet}
\end{figure}
Opposite to standard UNet, authors propose to use asymmetrically large encoder compared to the relatively thin decoder. ReLU activations are used as nonlinearities in this model. As a normalization layer, Group Normalization \cite{groupnorm} is employed due to the small batch size used during training. Group Normalization computations are independent of batch size, thus GN provides similar performance on small and large batch sizes. In the encoder part, feature maps are progressively downsampled by 2 and increased in number by the same factor. The number of residual blocks at each level equals to 1, 2, 2 and 4. In the decoder, however, the number of residual blocks remains equal to 1 across all levels. Authors propose to use additional Variational Auto-Encoder (VAE) branch for regularization. We follow this choice for training the ensemble model, but we decided against using it in the student model. The network architecture is illustrated in Fig.~\ref{fig:resunet}.\newline
We train this model with Adam optimizer using the same learning rate and learning schedule as for the previous baseline method. The loss function we used is the same as in the previous case. The patch size of 144x144x128 gave us better results compared to the smaller ones.
\subsection{Cascaded UNet}
\begin{figure}
\includegraphics[width=\textwidth]{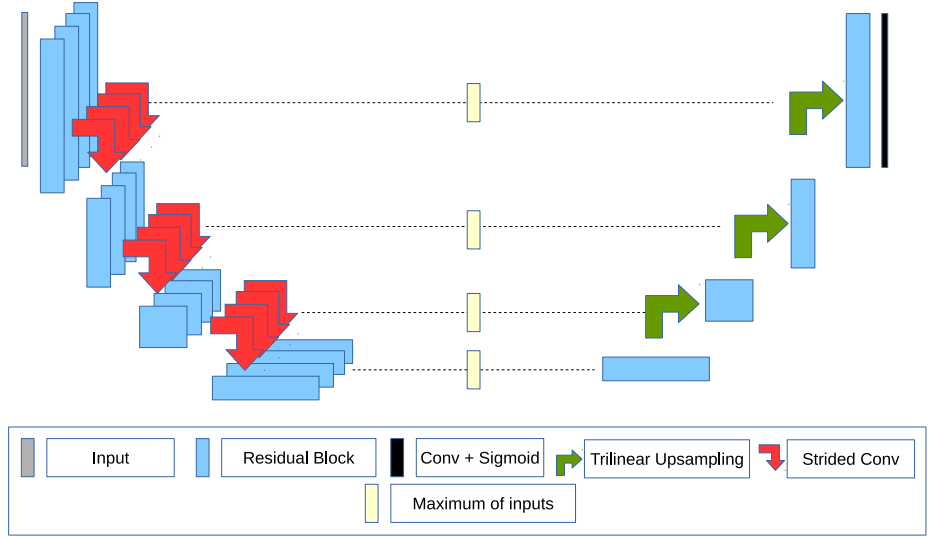}
\caption{Schematic representation of base network in the cascade. Each input modality has separate encoder. The encoder outputs are then merged together with elementwise maximum.}
\label{fig:cascunet}
\end{figure}

The third baseline method we use was previously introduced by the corresponding authors for participating in BraTS 2018. There we used a cascade of UNets \cite{lachinov1}, each one has multiple encoders that correspond to input modalities. In this method, we employed ReLU as the activation function. The network was built of basic pre-activation residual blocks that consist of two instance normalization layers, two ReLU activation layers and two convolutions with a kernel size of 3. We build a cascade of the models with the same topology that operate on the different scales of the input volume. Thus, we achieve an extremely large receptive field that can capture the global context. The architecture of base network in the cascade is illustrated in Fig.~\ref{fig:cascunet}.\newline
The model is trained with SGD with initial learning rate of 0.1, exponential learning rate decay with rate 0.99 for every epoch, the momentum of 0.9 and minibatch size equal to 4. All input samples were resampled to 128x128x128 resolution. The training was performed for 500 epochs.\newline
\section{Data preprocessing and augmentation}
As a preprocessing step, we normalize each input volume (modality) to have zero mean and unit variance for non-zero foreground voxels. This normalization is done independently for each input image.  For training Cascaded UNet we also resample input images to the resolution of 128x128x128.\newline
To enhance generalization capabilities of the networks we perform a large set of data augmentations during training. First, we crop random regions of the images in the way, that at least one non background pixel is present in the cropped fragment. Then we randomly scale, rotate and mirror these images across X and Y axes. We deciseded to keep the original orientation along Z axis. Finally, we apply intensity shift and contrast augmentations for each modality independently.\newline
\section{Knowledge Distillation}
We build the ensemble of above-named models by averaging the outputs. Next, we annotate all unlabeled data we have with this ensemble and use a combined set of manually and automatically labeled data as a training dataset. We pick architecture described in Section \ref{sec:resunet} without VAE branch as a student model and train it in the same way. Our experiments demonstrated that model with VAE under performs compared to the model without it. Due to increased dataset size we are no longer required to use regularization to get plausible results. For the submission to the evaluation system we trained the student model on all the data we have, with manual annotations and annotations from ensemble, except the dataset we are evaluating on.

\begin{figure}
\centering
\includegraphics[width=\textwidth]{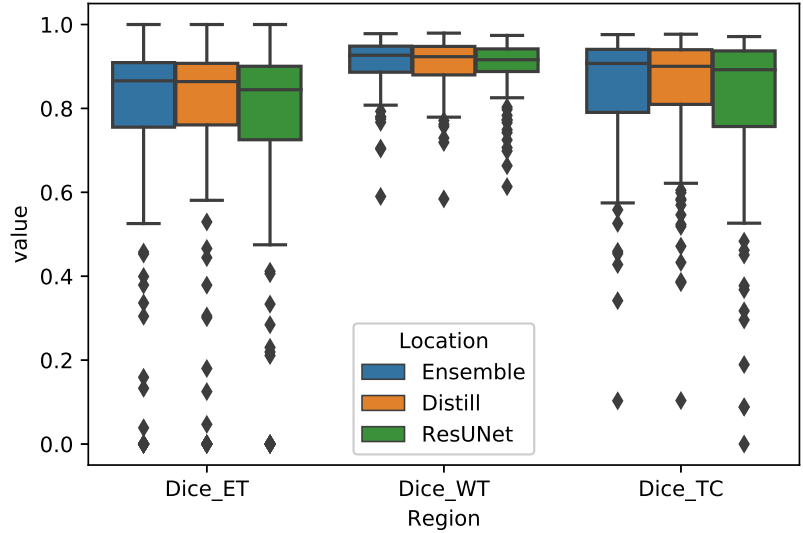}
\caption{The boxplot of the evaluation results on validation dataset.}
\label{fig:boxplot}
\end{figure}

\section{Results}

\begin{table}
\caption{Performance of the methods. Local validation scores are reported.}
\label{tab:local}
\centering
\begin{tabular}{|l|l|l|l|}
\hline
Method & Dice ET & Dice WT & Dice TC\\
\hline
UNet & 0.7836 & 0.9152 & 0.8743\\
Res UNet & 0.7392 & 0.9204 & 0.8754\\
Casc UNet & 0.9235 & 0.8925 & 0.8719\\
\hline
Distilled & 0.7440 & 0.9218 & 0.8835\\
\hline
\end{tabular}
\end{table}
\begin{table}
\caption{Performance of the methods. The scores were evaluated on validation dataset.}
\label{tab:validation}
\centering
\begin{tabular}{|l|l|l|l|}
\hline
Method & Dice ET & Dice WT & Dice TC\\
\hline
UNet & 0.7402 & 0.8974 & 0.8349\\
Res UNet & 0.7424 & 0.9018 & 0.8278\\
Casc UNet & 0.7307 & 0.8997 & 0.8335\\
Ensemble & 0.7562 & 0.9072 & 0.8435\\
\hline
Distilled & 0.7563 & 0.9045 & 0.8420\\
\hline
\end{tabular}
\end{table}

\begin{table}
\caption{Performance of the student model on testing dataset.}
\label{tab:test}
\centering
\begin{tabular}{|l|l|l|l|}
\hline
Method & Dice ET & Dice WT & Dice TC\\
\hline
Distilled & 0.7949 & 0.8874 & 0.8292\\
\hline
\end{tabular}
\end{table}
The training was performed with PyTorch Framework \cite{pytorch} and single Nvidia 2080Ti. Evaluation was performed locally using stratified train-test split, as well as using validation dataset and online evaluation platform. Due to sensitivity of validation metric to borderline cases (ex. empty mask), Cascaded UNet demonstrated high Dice score segmenting Enhancing Tumor by correctly predicting these few borderline cases in the validation split.\newline
On the validation dataset the ensemble of the models got the highest score. At the same time, the model trained with knowledge distillation scored almost the same as ensemble. As can be seen on graph \ref{fig:boxplot} and table \ref{tab:validation}, that represent results on validation dataset, student (distilled) model has lower Dice score variance of Enhancing tumor and Tumor core regions. On the local validation dataset (table \ref{tab:local}) the performance of the student model is comparable to the performance of the ensemble, except for ET class, where many empty masks are present. The final submission results of the distilled model on the test dataset are present in the table \ref{tab:test}. Surprisingly, the score for Enhancing Tumor got slightly increased compared to the validation results. That fact may indicate lower number of empty ET regions in test set compared to validation set.\newline
Performance of the student model across all the evaluation is comparable to the performance of the ensemble. Initially, we expected it to even surpass ensemble performance as it was demonstrated in \cite{distill2} due to larger number of samples to train on.
\begingroup
\let\clearpage\relax
\bibliographystyle{splncs04}
\bibliography{bibtex}

\begin{thebibliography}{10}
\providecommand{\url}[1]{\texttt{#1}}
\providecommand{\urlprefix}{URL }
\providecommand{\doi}[1]{https://doi.org/#1}

\bibitem{dataset2}
Bakas, S., Akbari, H., Sotiras, A., Bilello, M., Rozycki, M., Kirby, J.S.,
  Freymann, J.B., Farahani, K., Davatzikos, C.: Advancing the cancer genome
  atlas glioma mri collections with expert segmentation labels and radiomic
  features. Sci Data  \textbf{4},  170117 (Sep 2017).
  \doi{10.1038/sdata.2017.117},
  \url{http://www.ncbi.nlm.nih.gov/pmc/articles/PMC5685212/}, 28872634[pmid]

\bibitem{data1}
Bakas, S., Akbari, H., Sotiras, A., Bilello, M., Rozycki, M., Kirby, J.S.,
  Freymann, J.B., Farahani, K., Davatzikos, C.: Segmentation labels and
  radiomic features for the pre-operative scans of the tcga-gbm collection. The
  Cancer Imaging Archive  (2017). \doi{10.7937/K9/TCIA.2017.KLXWJJ1Q}

\bibitem{data2}
Bakas, S., Akbari, H., Sotiras, A., Bilello, M., Rozycki, M., Kirby, J.S.,
  Freymann, J.B., Farahani, K., Davatzikos, C.: Segmentation labels and
  radiomic features for the pre-operative scans of the tcga-lgg collection. The
  Cancer Imaging Archive  (2017). \doi{10.7937/K9/TCIA.2017.GJQ7R0EF}

\bibitem{dataset3}
Bakas, S., Reyes, M.M., Jakab, A., Bauer, S., Rempfler, M., Crimi, A.,
  Shinohara, R.T., Berger, C., Ha, S.M., Rozycki, M., Prastawa, M., Alberts,
  E., Lipkov{\'a}, J., Freymann, J.B., Kirby, J.S., Bilello, M.,
  Fathallah-Shaykh, H.M., Wiest, R., Kirschke, J., Wiestler, B., Colen, R.R.,
  Kotrotsou, A., LaMontagne, P., Marcus, D.S., Milchenko, M., Nazeri, A.,
  Weber, M., Mahajan, A., Baid, U., Kwon, D., Agarwal, M., Alam, M., Albiol,
  A., Albiol, A., Varghese, A., Tuan, T.A., Arbel, T., Avery, A., Pranjal, B.,
  Banerjee, S., Batchelder, T., Batmanghelich, N., Battistella, E., Bendszus,
  M., Benson, E., Bernal, J., Biros, G., Cabezas, M., Chandra, S., Chang, Y.J.,
  et~al.: Identifying the best machine learning algorithms for brain tumor
  segmentation, progression assessment, and overall survival prediction in the
  brats challenge. ArXiv  \textbf{abs/1811.02629} (2018)

\bibitem{3dunet}
{\c{C}}i{\c{c}}ek, {\"O}., Abdulkadir, A., Lienkamp, S.S., Brox, T.,
  Ronneberger, O.: 3d u-net: Learning dense volumetric segmentation from sparse
  annotation. In: Medical Image Computing and Computer-Assisted Intervention --
  MICCAI 2016. pp. 424--432. Springer International Publishing, Cham (2016)

\bibitem{resnet}
He, K., Zhang, X., Ren, S., Sun, J.: Deep residual learning for image
  recognition. In: 2016 IEEE Conference on Computer Vision and Pattern
  Recognition (CVPR). pp. 770--778 (June 2016). \doi{10.1109/CVPR.2016.90}

\bibitem{distill1}
{Hinton}, G., {Vinyals}, O., {Dean}, J.: {Distilling the Knowledge in a Neural
  Network}. arXiv e-prints arXiv:1503.02531 (Mar 2015)

\bibitem{batchnorm}
Ioffe, S., Szegedy, C.: Batch normalization: Accelerating deep network training
  by reducing internal covariate shift. CoRR  \textbf{abs/1502.03167} (2015),
  \url{http://arxiv.org/abs/1502.03167}

\bibitem{isensee2018}
Isensee, F., Kickingereder, P., Wick, W., Bendszus, M., Maier-Hein, K.H.: No
  new-net. In: Crimi, A., Bakas, S., Kuijf, H., Keyvan, F., Reyes, M., van
  Walsum, T. (eds.) Brainlesion: Glioma, Multiple Sclerosis, Stroke and
  Traumatic Brain Injuries. pp. 234--244. Springer International Publishing,
  Cham (2019)

\bibitem{lachinov1}
Lachinov, D., Vasiliev, E., Turlapov, V.: Glioma segmentation with cascaded
  unet. In: Crimi, A., Bakas, S., Kuijf, H., Keyvan, F., Reyes, M., van Walsum,
  T. (eds.) Brainlesion: Glioma, Multiple Sclerosis, Stroke and Traumatic Brain
  Injuries. pp. 189--198. Springer International Publishing, Cham (2019)

\bibitem{dataset1}
{Menze}, B.H., {Jakab}, A., {Bauer}, S., {Kalpathy-Cramer}, J., {Farahani}, K.,
  {Kirby}, J., {Burren}, Y., {Porz}, N., {Slotboom}, J., {Wiest}, R., {Lanczi},
  L., {Gerstner}, E., {Weber}, M., {Arbel}, T., {Avants}, B.B., {Ayache}, N.,
  {Buendia}, P., {Collins}, D.L., {Cordier}, N., {Corso}, J.J., {Criminisi},
  A., {Das}, T., {Delingette}, H., {Demiralp}, Ã., {Durst}, C.R., {Dojat}, M.,
  {Doyle}, S., {Festa}, J., {Forbes}, F., {Geremia}, E., {Glocker}, B.,
  {Golland}, P., {Guo}, X., {Hamamci}, A., {Iftekharuddin}, K.M., {Jena}, R.,
  {John}, N.M., {Konukoglu}, E., {Lashkari}, D., {Mariz}, J.A., {Meier}, R.,
  {Pereira}, S., {Precup}, D., {Price}, S.J., {Raviv}, T.R., {Reza}, S.M.S.,
  {Ryan}, M., {Sarikaya}, D., {Schwartz}, L., {Shin}, H., {Shotton}, J.,
  {Silva}, C.A., {Sousa}, N., {Subbanna}, N.K., {Szekely}, G., {Taylor}, T.J.,
  {Thomas}, O.M., {Tustison}, N.J., {Unal}, G., {Vasseur}, F., {Wintermark},
  M., {Ye}, D.H., {Zhao}, L., {Zhao}, B., {Zikic}, D., {Prastawa}, M., {Reyes},
  M., {Van Leemput}, K.: The multimodal brain tumor image segmentation
  benchmark (brats). IEEE Transactions on Medical Imaging  \textbf{34}(10),
  1993--2024 (Oct 2015). \doi{10.1109/TMI.2014.2377694}

\bibitem{myronenko}
Myronenko, A.: 3d {MRI} brain tumor segmentation using autoencoder
  regularization. CoRR  \textbf{abs/1810.11654} (2018),
  \url{http://arxiv.org/abs/1810.11654}

\bibitem{pytorch}
Paszke, A., Gross, S., Chintala, S., Chanan, G., Yang, E., DeVito, Z., Lin, Z.,
  Desmaison, A., Antiga, L., Lerer, A.: Automatic differentiation in {PyTorch}.
  In: NIPS Autodiff Workshop (2017)

\bibitem{distill2}
{Radosavovic}, I., {Dollar}, P., {Girshick}, R., {Gkioxari}, G., {He}, K.:
  {Data Distillation: Towards Omni-Supervised Learning}. arXiv e-prints
  arXiv:1712.04440 (Dec 2017)

\bibitem{unet}
Ronneberger, O., Fischer, P., Brox, T.: U-net: Convolutional networks for
  biomedical image segmentation. In: Medical Image Computing and
  Computer-Assisted Intervention -- MICCAI 2015. pp. 234--241. Springer
  International Publishing, Cham (2015)

\bibitem{instancenorm}
Ulyanov, D., Vedaldi, A., Lempitsky, V.S.: Instance normalization: The missing
  ingredient for fast stylization. CoRR  \textbf{abs/1607.08022} (2016),
  \url{http://arxiv.org/abs/1607.08022}

\bibitem{groupnorm}
Wu, Y., He, K.: Group normalization. CoRR  \textbf{abs/1803.08494} (2018),
  \url{http://arxiv.org/abs/1803.08494}

\end{thebibliography}
\endgroup
\end{document}